\documentclass[doublecol,figures,noheader]{epl2} 
\usepackage{epsf}
\usepackage{amsmath}
\usepackage{graphics,graphicx}
\usepackage{epstopdf}

\renewcommand{\revision}[1]{#1}

\title{\revision{Influence of the lattice topography on a three-dimensional, controllable Brownian motor}}\shorttitle{Influence of the lattice topography on a three-dimensional, controllable Brownian motor} 
\author{H. Hagman\thanks{E-mail: \email{henning.hagman@physics.umu.se}} \and C. M. Dion \and P. Sj\"{o}lund \and S. J. H. Petra  \and  A. Kastberg}
\shortauthor{H. Hagman \etal}
\institute{Department of Physics, Ume\aa\ University, SE-90187 Ume\aa , Sweden}
\pacs{32.80.Lg}{Mechanical effects of light on atoms, molecules, and ions}
\pacs{05.40.Jc}{Brownian motion}
\pacs{32.80.Pj}{Optical cooling of atoms; trapping}
\abstract{We study the \revision{influence of the lattice topography and the coupling between motion in different directions, for a}  three-dimensional Brownian motor based on cold atoms in a double optical lattice. Due to  controllable relative spatial phases between the lattices, our Brownian motor can induce drifts in arbitrary directions. \revision{Since the lattices couple the different directions, the relation between the phase shifts and the directionality of the induced drift is non trivial. Here is therefore this relation investigated}
\revision{experimentally by systematically varying the relative spatial phase in two dimensions, while monitoring the vertically induced drift and the temperature. A relative spatial phase range of $2\pi\times2\pi$ is covered.  We show that a drift, controllable both in speed and direction, can be achieved, \revision{by varying the phase both parallel and perpendicular to the direction of the measured induced drift.} The experimental results are qualitatively reproduced by numerical simulations of a simplified, classical model of the system}.}
\begin{document} 
\maketitle
\section{Introduction}
A Brownian motor (BM) converts random fluctuations into deterministic work  \cite{BM1 , BM2 , BM3}.  
Brownian motors exits naturally as, \emph{e.g.}, protein motors and intra-cell motion \cite{bioBM}, and a general understanding of their mechanism is of fundamental interest. The rectification mechanism requires that the system is both brought out of thermal equilibrium \cite{BM1} and spatially or temporally asymmetric \cite{Curie , BM1}. Although it has not been proven, fulfilling these two requirements are generally sufficient to realise a BM.
These requirements can be met using ultra cold atoms trapped in optical lattices \cite{lin-p-lin1 , OL}. Generally, the symmetry is broken by using a spatially asymmetric or flashing potential. In this type of noise rectifier, the direction of the induced drift is often fixed for a given potential or controllable in just 1D \cite{BM1 , BM2 , BM3 , r4 , RenzoniBM , RenzoniBM2 , RenzoniBM3 , GrynbergBM}. 

Our BM \cite{OurBM1 , newBM} is based on two symmetric potentials, with an asymmetry that originates from a combination of a relative spatial phase  shift between the potentials and an unequal transfer rate between them. Our system possesses an inherent ability to induce drifts in an arbitrary direction in three dimensions for the ultra cold atoms interacting with the potentials. The direction is mainly controlled by the relative spatial phase between the two potentials. \revision{In \cite{OurBM1 , newBM} we demonstrated that our BM does work in more} \revision{than one dimension, but how this works was not investigated. This question is especially complicated since the four-beam configuration induces a coupling between the different spatial dimensions. 
This coupling affects the 3D behaviour of our BM profoundly and is the key to any understanding and control of the three-dimensional aspects of our BM.} 

\revision{In this paper, we present a study of the dimensional coupling between the relative spatial phases and its influence on the induced drift, in two dimensions}. This is done both experimentally and with numerical simulations.  \revision{This study results in a better understanding of the multidimensionality of our BM, and in a good control to set the induced drift to a chosen speed and direction.} 
This also renders a controlled dynamical BM possible, where a time-dependent phase can  induce realtime drifts along any pre-chosen trajectory. \revision{It even allows to control the induced drift in the vertical direction by purely controlling the phase shift horizontally, and vice versa.}
By studying the phase dependence, we can also study the coupling between the dimensions of the two potentials. Information about what potentials the atoms really experience and the adiabaticy \cite{OL} of the interaction can therefore be investigated. \revision{The numerical simulations are done with a simplified, classical  model, which mimics the main features of our BM. This  qualitatively reproduce the main features of the results from the experiments.}
\section{System}
The potentials used are realised from a double optical lattice (DOL) \cite{DOL , setups}, interacting with $^{133}$Cs. 
An optical lattice is a periodic potential where atoms can be cooled and trapped due to dissipation and light shifts \cite{lin-p-lin1}, and is created from the interference between two or more light fields. A DOL consists of two spatially overlapped state-dependent optical lattices \cite{DOL , setups}.

\begin{figure}[htbp]
\centering
\includegraphics[width=8cm]{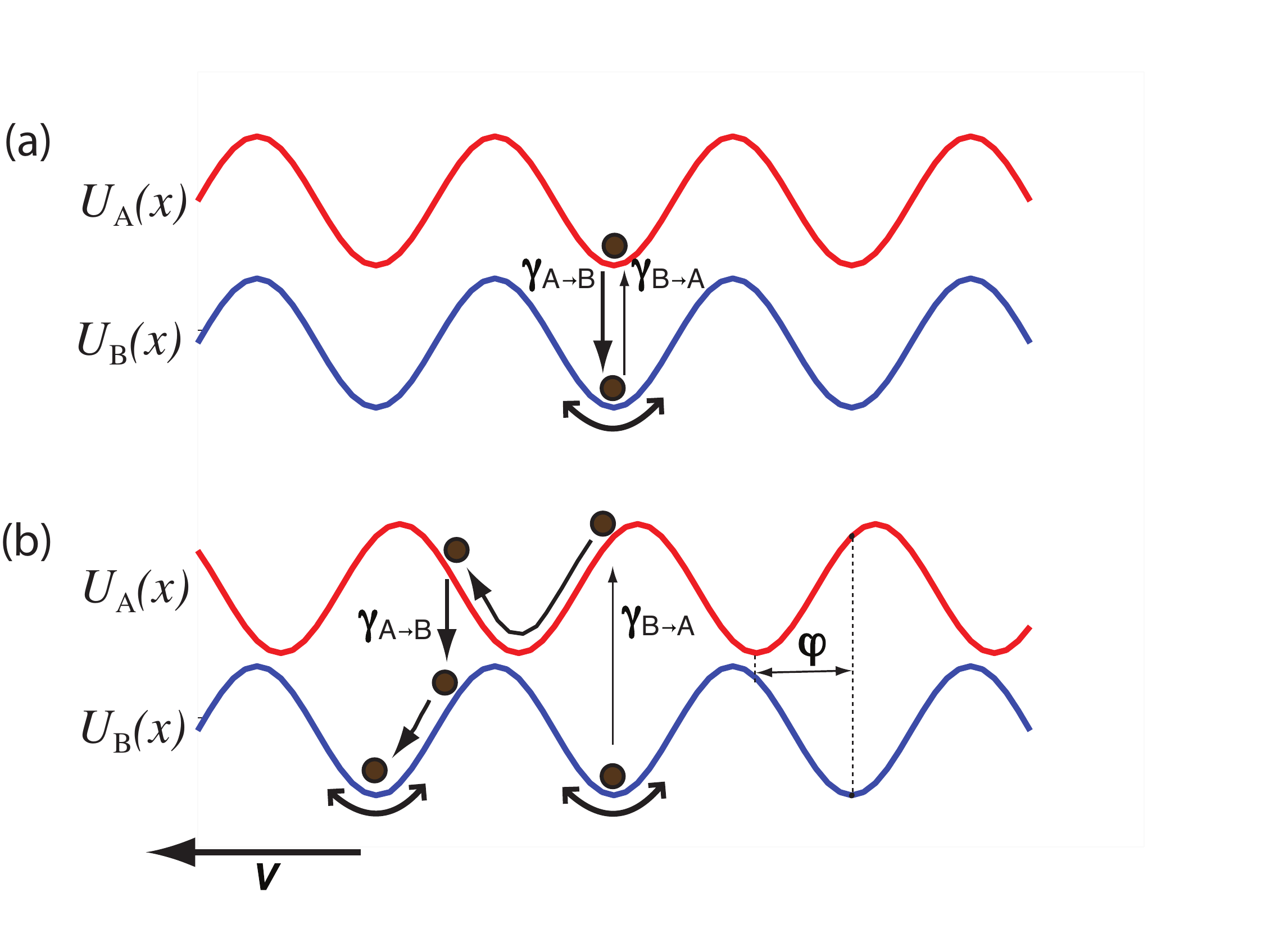}
\caption{(Colour online) Simplified, classical model of the BM mechanism. (a) $\varphi = 0$, the system is spatially symmetric and the particle undergoes Hamiltonian oscillation at the bottom of the wells. No drift is induced. (b) $\varphi \ne  0$, the spatial-temporal symmetry is broken and a drift $v$ to the left is induced. \revision{In both cases there is a large difference in the transfer rates $\gamma$ between the lattices,  indicated with arrow of different thickness in the figure. Note that this a strongly simplified model, presented to provide an understanding of the phase dependence. Features such as diffusion and friction,  vital for the BM, are not present in this model}.}
\label{idea}
\end{figure}
To qualitatively understand the induced drift dependence on the relative spatial phase, a simple classical model is used \cite{BMsim , ratide}, see figure \ref{idea}.
Consider a classical Brownian particle situated  in either of two symmetric and periodic potentials \revision{($U_\mathrm{A}$ and $U_\mathrm{B}$)}, coupled with unequal transfer rates ($\gamma_\mathrm{A\rightarrow B}$ and $\gamma_\mathrm{B\rightarrow A}$). The particle will shift back and forth between the potentials, but on average spend longer time in one of them. When the potentials are in phase, the trapped particle will jump between potentials and undergo Hamiltonian oscillations near the bottom of them, see figure \ref{idea}a. If the potentials  are given a non-zero relative spatial phase shift, see figure \ref{idea}b, with transfer rates properly chosen, the particle will spend long enough time in the long-lived potential in order to be located close to it{s} minimum. After a certain time the particle is transferred to the short-lived potential, where it will be located at a slope of the potential. After a short while, it will be pumped back to the long-lived potential. The particle will therefore, on average gain kinetic energy in a certain direction when jumping between the potentials \cite{ratide}. \revision{The} \revision{optimal ratio between the transfer rates has been experimentally determined to $\gamma_\mathrm{A\rightarrow B}: \gamma_\mathrm{B\rightarrow A} \simeq 9:1$ \cite{newBM}. This is also related to the time atoms spend in each potential. The time spent in potential A should optimally be of the order of  the the inverse of this potential's oscillation frequency \cite{phd peder}, while the optimal time spent in potential B is much larger. The oscillation frequencies are typically of the order of 100 kHz.}
\section{Methodology}
Detailed descriptions of the general experimental setup are given in \cite{Johan2002 , Harald2002 , setups , DOL}. More details concerning the construction and the control of the BM are found in \cite{OurBM1 ,newBM , balance}. In short, cesium atoms are first accumulated in a magneto-optical trap, from where they are transfered to the double optical lattice. 
Both lattices (A and B) have a 3D topography and are constructed from  four beam configurations, in which two beams are propagating in the \emph{xz}-plane, and two in \emph{yz}-plane. All beams have a $45\,^{\circ}$ angle with respect to the vertical \emph{z}-axis with polarizations perpendicular to the plane of propagation ({``3D lin$\perp$lin configuration'' \cite{lin-p-lin1}}). The lattices are state-dependent and operate from different hyperfine ground states within the Cs D2 line (6s $^2$S$_{1/2}$ $\rightarrow$ 6p $^2$P$_{3/2}$). Lattice A operates close to the $F_\mathrm g = 3 \rightarrow F_\mathrm e = 4$ transition and lattice B operates close to the $F_\mathrm g = 4 \rightarrow F_\mathrm e = 5$ transition. 
The irradiances and the frequencies of the four beams are chosen to generate pumping rates between the lattices that maximise the induced drift.

The optical path lengths are controllable in all beams, resulting in a control of the relative spatial phase. This dependence originates from a slight difference in wavelength between the lattices. The difference is small enough for the relative spatial phase to be effectively the same throughout the whole atomic cloud \cite{DOL , balance}.

The induced drift velocity is obtained by a ballistic time-of-flight
method (TOF), where the trapped atoms are released by quickly turning
off the lattice beams. The atoms then freely fall through a resonant
probe located approximately 5 cm below the trapping region. The
fluorescence from the atoms passing through the probe is detected and
gives a measurement of the arrival time $t$ of the atomic cloud
\cite{TOF , sisyphus}. Since the drift velocity is constant
\cite{OurBM1}, the arrival time can be converted into drift velocity
through
\begin{equation}
v_z = \frac{gt^2 - 2s_\mathrm{p}}{2(t+\tau)},
\end{equation}
where $g$ is the gravitational constant, $\tau$ is the interaction time of the atoms with the DOL and $s_\mathrm{p}$, is the distance to the probe.
\section{Results}
\begin{figure}[htbp]
\centering
\includegraphics[width=8.7cm]{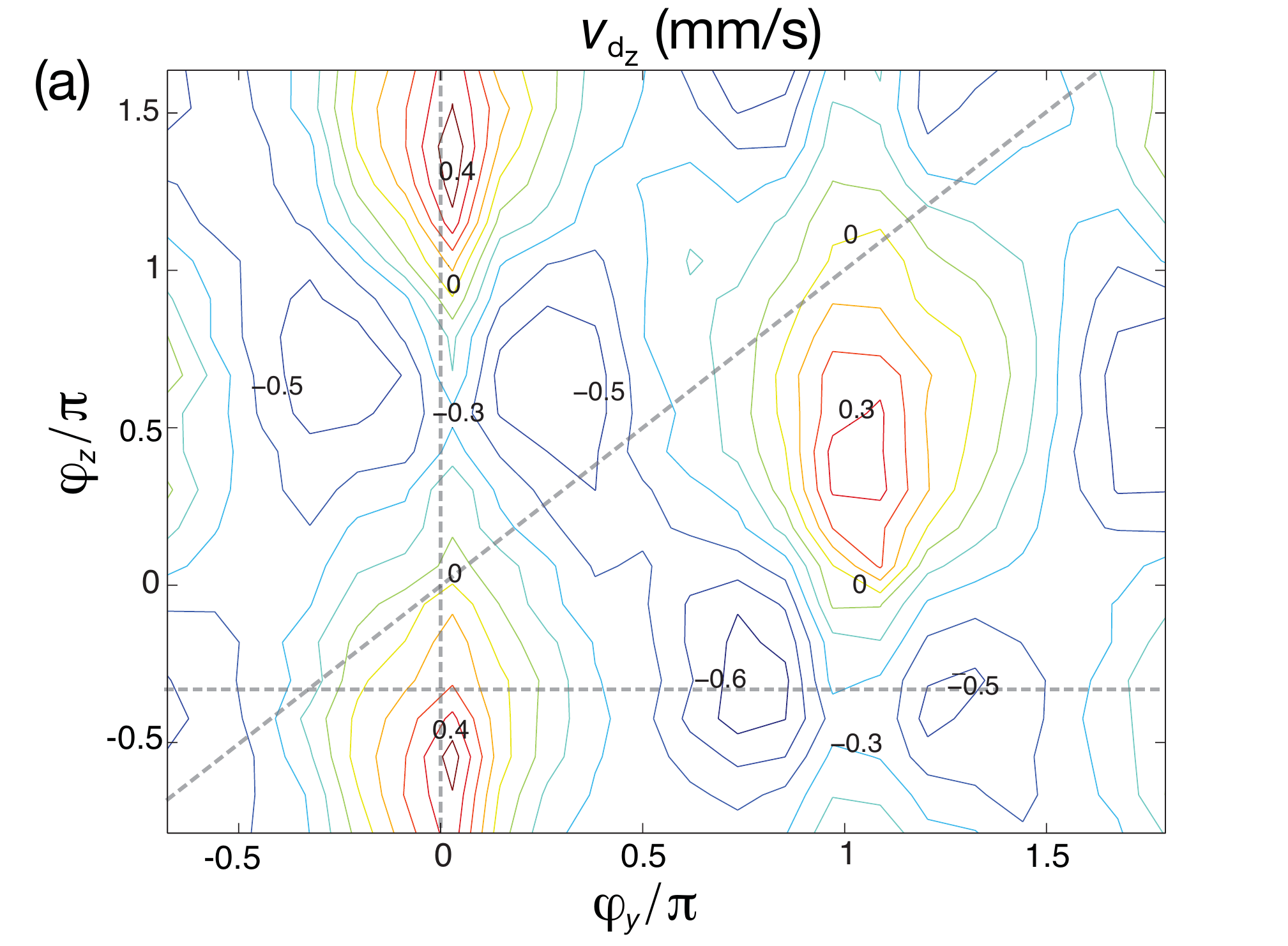}
\\
\centering
\includegraphics[width=8.7cm]{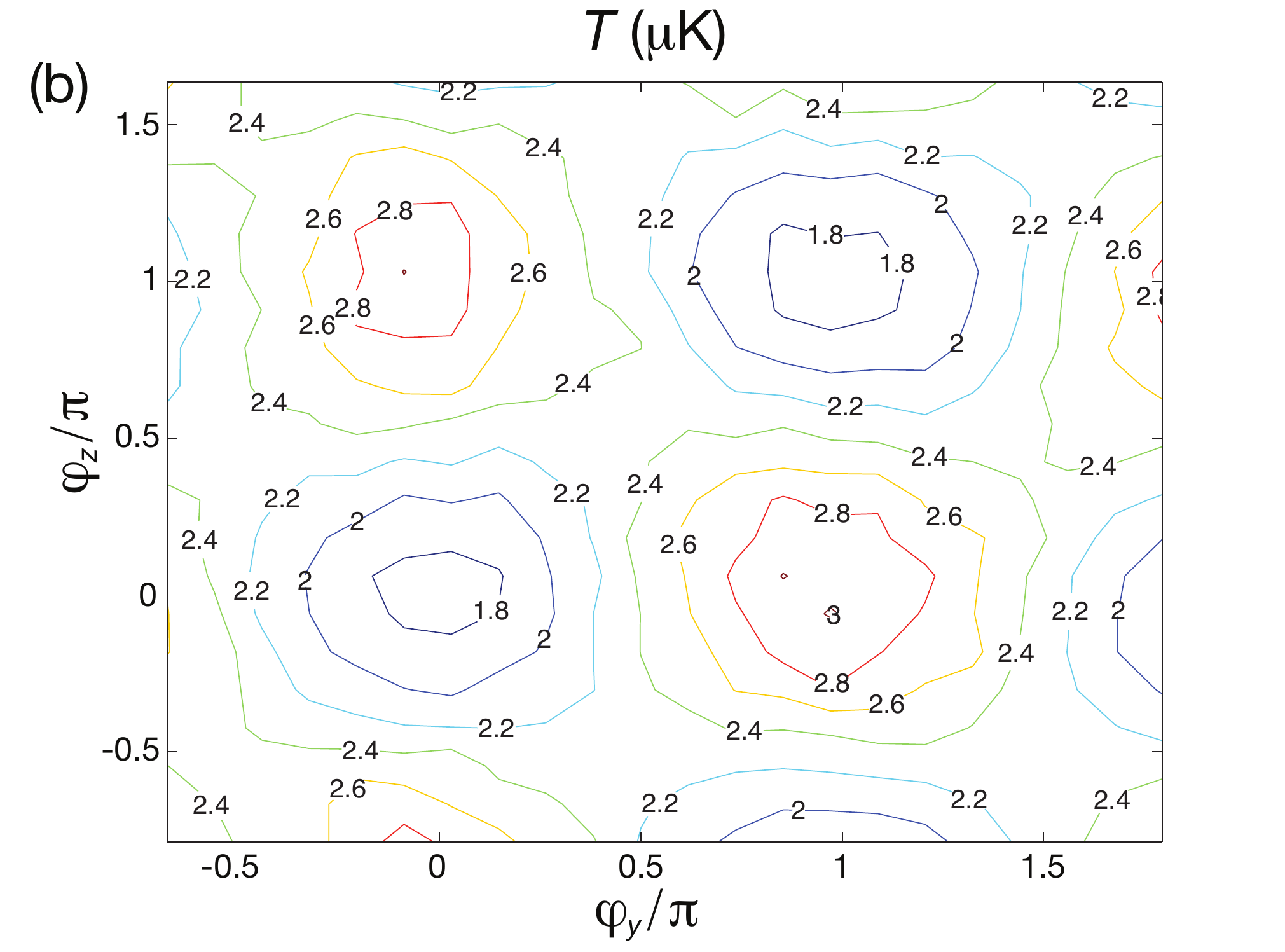}
\caption{(Colour online). (a) Measured induced drift, and (b) temperature, as a function of the relative spatial phase along the \emph{y}- and \emph{z}-axes. A difference in the pattern created by the induced drift maxima and minima is evident. The dashed lines indicates the cuts shown in figure \ref{cuts}.}
\label{2dratchet}
\end{figure}

To investigate the relative spatial phase dependency experimentally, we change the phase along \emph z ($\varphi_z$) and \emph y ($\varphi_y$) by slightly more than 2$\pi$. This is done in about 20 steps in each direction,  covering all steps in \emph{z} for each \emph{y} value. Here, $\varphi_{z,y} = 0$ corresponds to a perfect overlap of the two lattices. Since these intervals exceed $2\pi$,  whole periods along both the \emph{z} and \emph{y} axes are covered.
For each combination phases, the TOF signal is measured. To improve statistics, an average over five TOF measurements is made for each combination. A compilation of  the measurements is shown in figure \ref{2dratchet}. Also displayed is the temperature measured at each step, which is used to determine the phase.

The origin, where $\varphi_{y,z}$ is zero, is chosen at a minimum of the temperature, since a perfect spatial overlap of the lattices is assumed to correspond to a temperature minimum \cite{DOL , setups}. The scale is obtained by measuring the periodicity of the temperature and comparing it to the change of the optical path lengths. The measurement indicates that the temperature maxima occur when the diabatic potentials \cite{lin-p-lin1 , OL} are shifted with a relative spatial phase of $\pi$. Since the induced drift velocities and the temperature are measured simultaneously, the relative phase scale obtained from the temperature is applied to the BM.

Throughout the measurement, lattice B is red-detuned below resonance by $40\Gamma$ and has an irradiance of 0.2 mW/cm$^2$ in each beam and lattice A is red-detuned by $33\Gamma$ and has an irradiance of 0.7 mW/cm$^2$ in each beam, where  $\Gamma$ is the natural line width of the excited state ($\Gamma/2\pi = 5.2$ MHz). 
These parameter values are chosen to optimize the induced drift \cite{newBM} while still having a strong TOF signal with a good signal to noise ratio.  

\begin{figure}[htbp]
\centering
\includegraphics[width=8cm]{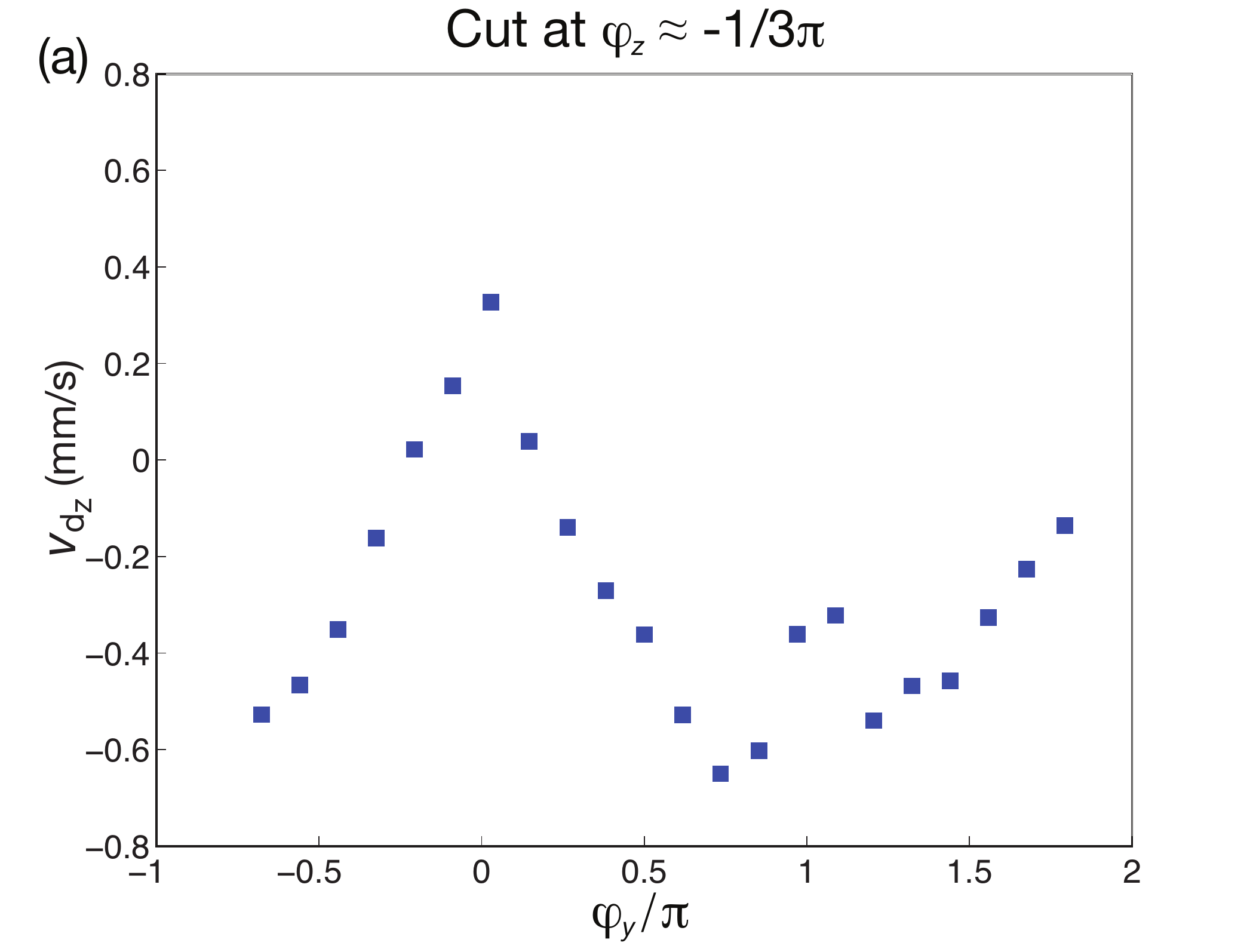}
\includegraphics[width=8cm]{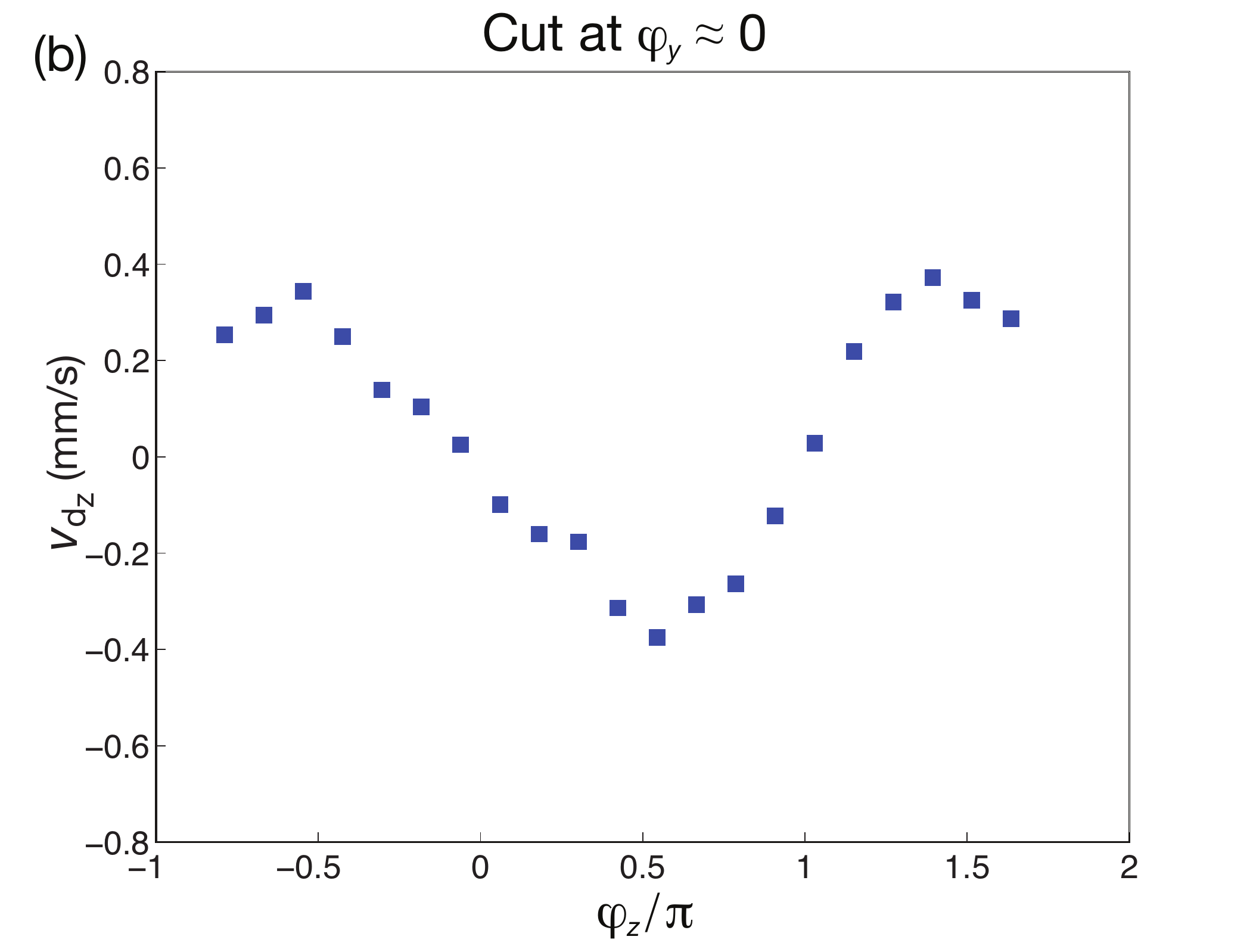}
\includegraphics[width=8cm]{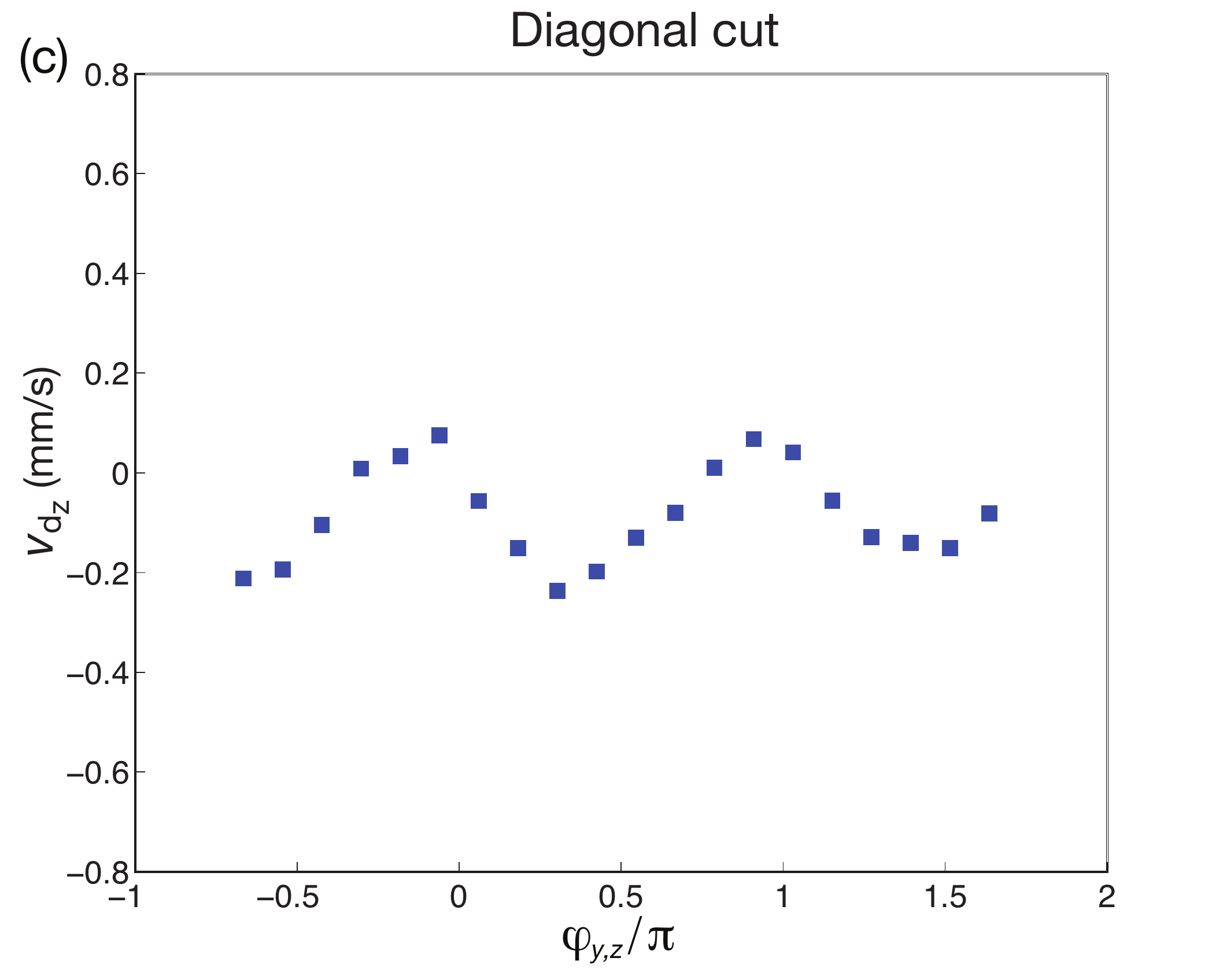}
\caption{Cuts in the two dimensional plot of figure \ref{2dratchet}a. (a) Cut along \emph{y}-axis, at $\varphi_z \simeq \frac{1}{3}\pi$. A clear splitting of the minima is visible at $\varphi_y \simeq \pi$ (b) Cut along \emph{z}-axis, at $\varphi_y \simeq 0$. The induced drift shows a close to sinusoidal shape.  (c) A diagonal cut, at $\varphi_z \simeq \varphi_y$. A doubled frequency is evident.}
\label{cuts}
\end{figure}
Even though the induced drift has the same periodicity, the topography along the \emph{z} and \emph{y} axes differ. Along the \emph{z} axis, the drift has a close to sinusoidal dependence on $\varphi_z$, for any choice of $\varphi_y$ ({see figure \ref{cuts}b), although the amplitude and the phase of the drift velocity strongly depend on the relative spatial phase along the \emph{y} axis. 

Along the \emph y axis, however, the induced drift clearly deviates from a sinusoidal shape, having two minima per period (see fig \ref{cuts}a). \revision{Such a difference between the \emph z and \emph y intersections is expected since the lattice structure differs slightly along these directions \cite{lin-p-lin1}}.
By moving diagonally in the \emph{yz} phase plane, the periodicity is doubled (see fig \ref{cuts}(c)).
In order to understand the qualitative behaviour of our Brownian motor, we have performed simulations, using a simplified, \revision{classical} model of our system \cite{BMsim , ratchet2 , ratide}.
The two-dimensional numerical simulations are done for a classical Brownian particle situated in either of two potentials, corresponding to the lowest diabatic potential \cite{BMsim , DOL} of each lattice in a DOL configuration, identical to the one used in the experiments. The dynamics of the atom is obtained from the Fokker-Planck equation \cite{Fokker1 , Fokker2}.
The induced drifts in the \emph z and \emph y directions as a function of the relative spatial phase are shown in figure \ref{sim}. 
\begin{figure}[htbp]
\centering
\includegraphics[width=8cm]{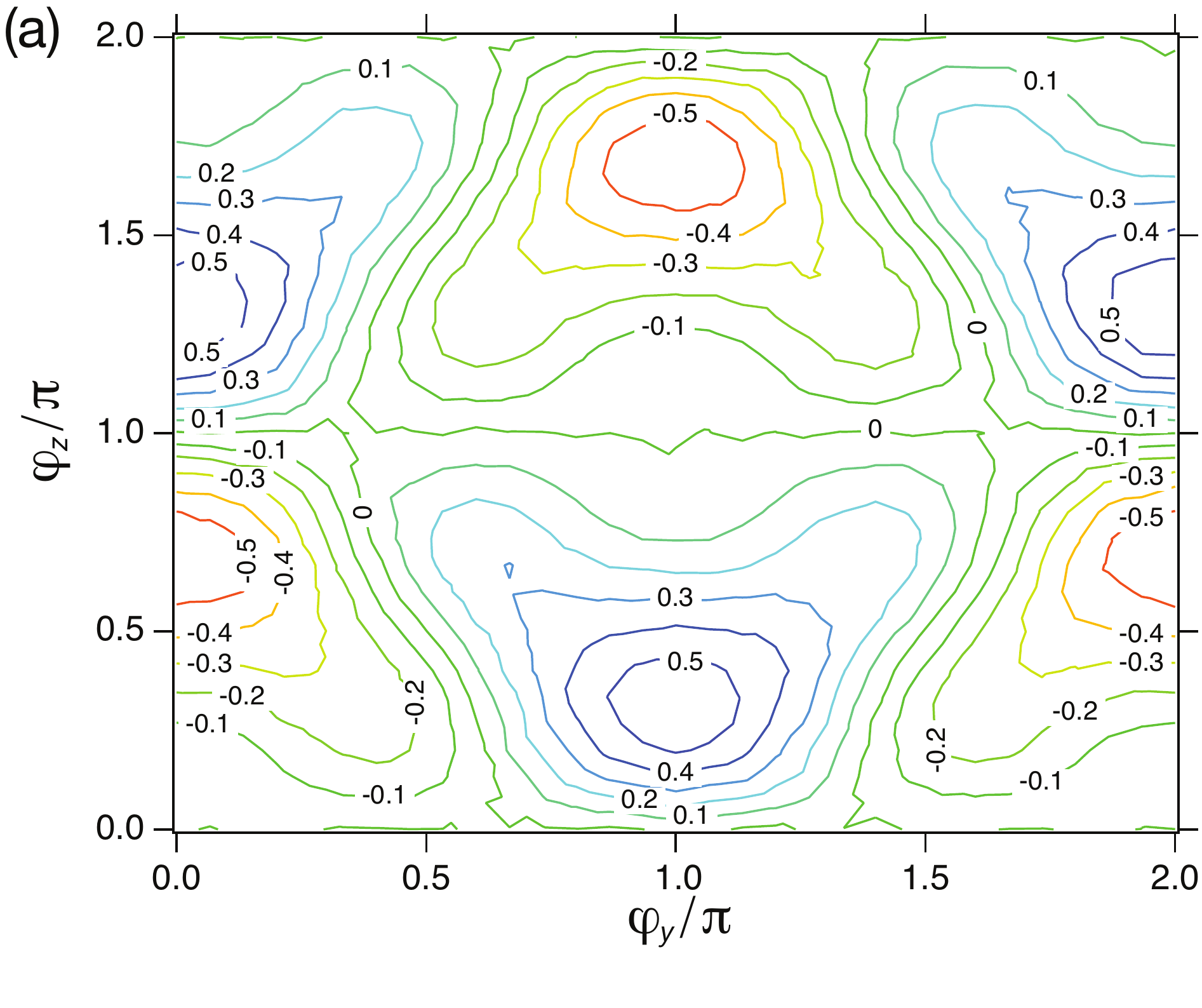}
\includegraphics[width=8cm]{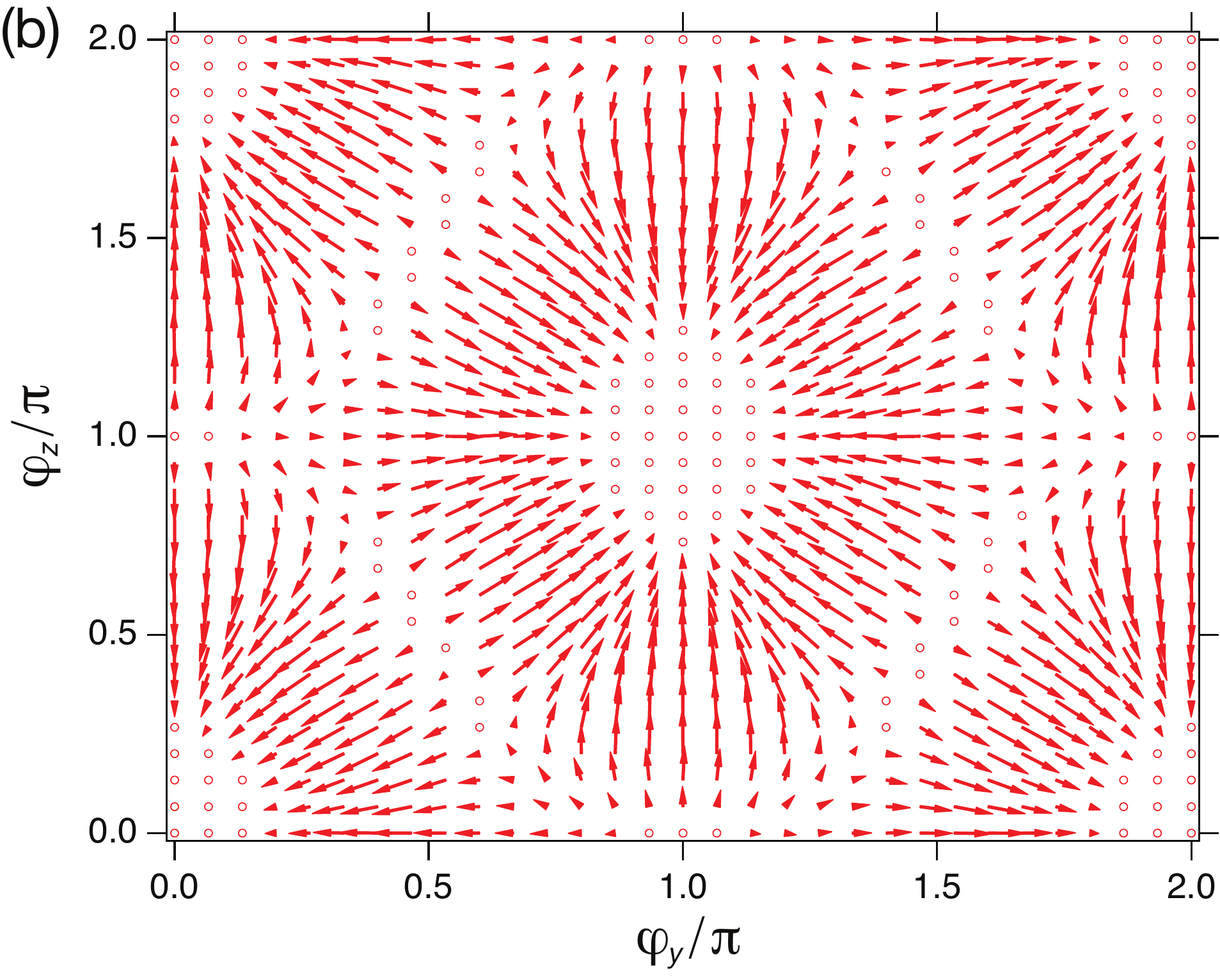}
\caption{(Colour online) Simulation of the induced drift as a function of the relative spatial phases $\varphi_z$ and $\varphi_y$.  The speed is given in mm/s. (a) False colour contour  plot of induced drift in \emph z. (b) Arrow plot of the induced drift in \emph z and \emph y. The arrows indicate the size and the direction of the induced drift. \revision{The phase at wich the velocity reaches its maximum differs along the \emph z and \emph y directions. 
This agrees with the result obtained in \cite{newBM}.}} 
\label{sim}
\end{figure}
The simulations show a clear periodic behaviour in both the \emph z and \emph y directions, and the fact that the induced drift maxima in positive and negative directions do not lie on a straight  line along the \emph y axis  shows that the potentials in different directions are coupled. \revision{That is, it is not possible to go from a maxima to a minima by purely changing $\varphi_y$,  $\varphi_z$} must vary as well.  This is also visible in the equation describing the potentials,
\begin{eqnarray}
\lefteqn{U_\pm = \frac{4\hbar\Delta^{\prime}_0}{45}\{23[\cos^2(k_x x) + \cos^2(k_y y) ]  {} }
\nonumber\\
& & {}\mp  44\cos(k_x x)\cos(k_y y)\cos(k_z z)\},
\end{eqnarray}
where $\Delta_0^{\prime}$ is the scattering rate and \emph{k} is the effective angular wave vectors along the axes \cite{lin-p-lin1}.
This equation also shows that the potentials are identical in \emph x and \emph y, which would give identical relative spatial phase dependencies in \emph x and \emph y.
The simulations agree with the measurement, where a clear spatial coupling between the potentials is evident. Both also show a clear pattern, and even though the pattern does differ, this confirms that we can control our BM via the phase in arbitrary directions. \revision{While an extension of the simulations to 3D  is conceptually straightforward, it would be computationally demanding. The third dimension would, if anything, just enhance the coupling observed in 2D. Moreover, while this model contains the basic ingredients of our BM (diffusion, friction, coupled potentials, etc.), it neglects potentially important factors (such as the spatial dependence of diffusion and friction or the presence of a manifold of potentials due to the magnetic substates), so no far reaching conclusions should be drawn from the comparison of the simulations with the experimental results.}
\section{Discussion}
The induced drift of our BM can be controlled by the difference between the transfer rates between the lattices, the diffusion, the potential height and the relative spatial phase between the lattices. The dependence of the first three parameters is investigated in \cite{newBM , OurBM1 , ratchet2}. The phase dependency was there fully investigated for drifts measured along only in one dimension, \revision{and phases varied along the same direction. Drifts }\revision{in more than one dimension were demonstrated but drifts induced by a phase shift in a perpendicular direction were not investigated at all. Since the potentials couples the different directions, the non-trivial relation between the phase shift of the lattices and the directionality of the induced drift is at the heart of the multidimensionality of our BM. Such an investigation is therefore of fundamental interest and} \revision{necessary obtain any understanding or control of our BM.}

By surveying the vertically induced drift by $2\pi\times 2\pi$ in the \emph{yz} phase plane, \revision{the influence of the lattice topography on} the structure and periodicity of the drift can be determined. Due to the identical potentials in the \emph x and \emph y directions, a generalisation to three dimensions is possible. The induced drift dependence on the transfer rates, \emph{i.e.}, the irradiances and detunings, has earlier been investigated \cite{newBM , OurBM1}, and thus we can now fully describe our three-dimensional BM, and with good control set the induced drift velocity to any arbitrary direction, with a controllable magnitude in three dimensions. Such control is necessary in future experiments where fast phase shifts should render it possible to realise a dynamical Brownian motor, where the atoms can change direction in real time and therefore travel along more or less any pre-chosen trajectory, limited only by the lifetime of the DOL-caused diffusion. These fast phase shifts can be realised with electro-optical modulators (EOM). The implementation of these EOMs also makes it possible to scan the phase over a range of 2$\pi$ in all three dimensions simultaneously, which  is not possible with the current set-up. 

The multidimensional \revision{coupling between the} phase dependency \revision{of the lattice topography and the induced drift}  render it possible to redirect the motion in a certain direction without changing the phase in that direction. The induced drift can, by changing the phase in an orthogonal direction, be enhanced, put to zero or even be inverted. \revision{This is clearly evident in figure \ref{2dratchet}(a), and has not been shown before.}

The measured periodicity of the temperature also confirms that the diabatic potentials are the relevant potentials in the system, as shown in \cite{newBM}. 
The two potentials in the classical simulations are therefore chosen as the two lowest diabatic potentials of \revision{the two} state-dependent potentials of the DOL used in the experiment \cite{BMsim}. The simulation
qualitatively reproduces the main features of our BM although it is based on a classical model. A closer comparison with the experimental result show clear differences, especially the location of the maxima in induced drift velocity in the \emph z direction. This shows that the simple classical picture is inadequate for extracting more detailed information about our system, and that not just the lowest diabatic potential is relevant. In reality an optical lattice works on a manifold of potentials, each corresponding to a transition between different magnetic sub-states \cite{OL}.
This manifold of potentials is also of importance in the model explaining the relative spatial phase dependence of the temperature in a DOL \cite{setups}, seen in figure \ref{2dratchet}b. The dependence of the temperature on the phase also affects our BM, since a change in temperature reflects different diffusion and friction in the system. This coupling may indeed contribute to the double peak structure of the maxima in the induced drift in the negative \emph z-direction, since the splitting overlaps with the temperature maxima. Moreover, the phase in the \emph x direction is  roughly set to zero at the beginning of the experiment by measuring the minimum in temperature. During the scan this value may change slightly but the effect is small, as long as the phase is small. Finally, a full physical understanding of the structure of our BM calls for a quantum model where the entire manifold of potentials that exist in the real DOL is accounted for. Such a model is under construction.
\section{Conclusion}
In summary, \revision{we have shown that the relative spatial phase of the lattices, due the the four beam configuration of the latter,  couple dimensionally the induced drift. This affects the multidimensional behaviour of our Brownian motor profoundly. These effects have been investigated and we showed that the vertical induced drift can be controlled by either of the vertical or the horizontal phase, or a combination of the two.}
This was done by measuring the vertical drift velocity in a $2\pi\times 2\pi$ area in the \emph{yz} phase plane. From the measurement, the structure and periodicity of the drift was extracted. This gives us full control over our three dimensional Brownian motor.
\acknowledgments
We thank S. Jonsell for helpful discussions. This work was supported by Knut och Alice Wallenbergs stiftelse, Vetenskapsr\aa det / Swedish research council, Carl Tryggers stiftelse and Kempestiftelserna. A part of this research was conducted using the resources of the High Preformence Computing Centre North (HPC2N).
\bibliographystyle{prsty}
\bibliography{bibfiles/shortnames,bibfiles/references}
\end{document}